\documentclass[letterpaper,journal]{IEEEtran}
\usepackage{amsmath,amsfonts}
\usepackage{array}
\usepackage[caption=false,font=normalsize,labelfont=sf,textfont=sf]{subfig}
\usepackage{textcomp}
\usepackage{stfloats}
\usepackage{url}
\usepackage{graphicx}
\usepackage{cite}
\usepackage{multirow}
\usepackage{booktabs}
\usepackage{needspace}
\usepackage{microtype}

\begin{document}

\title{Graph Learning for Cross-Subject, Cross-Population EEG Emotion Decoding and Model-Derived Spatial-Spectral Neural Signatures}

\author{Dongyi He, Bin Jiang, Xiangkai Wang, Yun Zhao, Gen Li, Hongjie Yan, Wai Ting Siok, and Nizhuan Wang%
\thanks{This research was supported by the Natural Science Foundation of Chongqing under Grants CSTB2025NSCQ-JM002, CSTB2025NSCQ-GPX0794, and CSTB2024NSCQ-MSX0118; the Scientific and Technological Research Program of the Chongqing Education Commission under Grants KJZD-K202303103, KJZD-K202501107, and KJQN202501104; the Chongqing Municipal Key Project for Technology Innovation and Application Development under Grants CSTB2024TIAD-KPX0042 and CSTB2025TIAD-KPX0002; The Hong Kong Polytechnic University under Internal Grant P0048377, Departmental Collaborative Research Fund Grant P0056428, and Collaborative Research with World-leading Research Groups Fund Grant P0058097; and the Research Grants Council Collaborative Research Fund under Grant C5033-24G.}%
\thanks{Dongyi He, Bin Jiang, and Xiangkai Wang are with the School of Artificial Intelligence, Chongqing University of Technology, Chongqing 400054, China. Dongyi He, Wai Ting Siok, and Nizhuan Wang are with the Department of Language Science and Technology, The Hong Kong Polytechnic University, Hung Hom 999077, Hong Kong SAR, China. Yun Zhao is with the School of Smart Health, Chongqing Polytechnic University of Electronic Technology, Chongqing 401331, China. Gen Li is with the School of Pharmacy and Bioengineering, Chongqing University of Technology, Chongqing 400054, China. Hongjie Yan is with the Affiliated Lianyungang Hospital of Xuzhou Medical University, Lianyungang 222002, China.}
\thanks{Corresponding authors: Bin Jiang (e-mail: jb20200132@cqut.edu.cn) and Nizhuan Wang (e-mail: wangnizhuan1120@gmail.com).}
}

\maketitle

\begin{abstract}
Cross-subject emotion decoding from electroencephalography (EEG) requires representations that accommodate individual variability while preserving spatial--spectral structure for interpretation. This study introduces EmoDiPyraTrans, a differential graph Transformer that integrates adaptive graph recurrence, differential attention, pyramid fusion and distribution regularization over sequential relative power spectral density graphs. Across SEED, FACED, MAHNOB-HCI, DEAP and DREAMER, the model achieved the highest participant-mean accuracy and positive-class F1 among the evaluated methods, with accuracy and F1 both reaching $0.928$ on SEED. On DEP-EEG, positive-versus-neutral accuracy reached $0.802$ within healthy controls and $0.704$ within participants with depression, compared with $0.591$ under healthy-to-depression transfer and $0.581$ with mixed-population development. Complementary SEED analyses identified distributed spatial weighting and an alpha-centred spectral preference, while configurations averaging six channels retained near-full performance. These findings link generalization assessment with model-derived candidate signatures to support interpretable EEG emotion decoding, with code available at \url{https://github.com/hdy6438/EmoDiPyraTrans}.
\end{abstract}


\section{Introduction}
\IEEEPARstart{E}{lectroencephalography} (EEG)-based emotion decoding infers affective states from non-invasive recordings of brain activity, providing a basis for affective brain--computer interfaces whose application to new users depends on cross-subject generalization \cite{huang2021neurofeedback}. Because individual anatomy, skull conductivity and electrode conditions influence EEG measurements \cite{antonakakis2020inter,kappenman2010effects}, prediction across participants requires representations that accommodate variation arising from these factors. At the same time, emotion-relevant information spans multiple channels and frequency bands \cite{zheng2015investigating}, motivating frameworks in which cross-subject representation learning is coupled with analysis of the spatial and spectral features supporting prediction.

To capture this spatiotemporal structure, multiscale convolutions are used in TSception to learn temporal dynamics and hemispheric asymmetry \cite{ding2022tsception}, while dynamical graph convolutional networks learn relationships among EEG channels \cite{song2018eeg}. Graph-based representations are further integrated with Transformer modelling in EmT, through which inter-channel structure and temporal dependencies are jointly considered for cross-subject emotion recognition \cite{ding2025emt}. In parallel, multisource adaptation methods use unlabelled target recordings to reduce discrepancies between source and target representations \cite{chen2021ms,wang2024multi}, complementing architectural modelling with distribution alignment. Together, these developments motivate integrating spatiotemporal representations and distribution regularization with channel-level readouts, so that predictive performance can be examined alongside the spatial--spectral features learned by the model.

As these modelling strategies are extended to cohorts with different clinical characteristics, evaluation across populations becomes central to characterizing how decoding performance varies with the composition of the development data. In DEP-EEG, positive-versus-neutral recordings from healthy controls and participants with depression \cite{huang2021neurofeedback,guan2025method} support complementary evaluations in which within-population testing assesses prediction in new participants from the same cohort and healthy-to-depression transfer evaluates performance across clinical groups. Mixed-population development extends these comparisons by examining how the combination of healthy and depression data influences emotion decoding in participants with depression, thereby characterizing performance across development populations and clinical settings.

Alongside evaluation across participants and populations, spatial and spectral analyses connect decoding performance to the features supporting prediction, as illustrated by studies of critical channels and frequency bands \cite{zheng2015investigating}. Related work in ERTNet examines learned temporal and spatial filters \cite{liu2024ertnet}, providing an additional basis for linking model representations to emotion decoding. Building on these approaches, attention maps can be combined with channel pruning and retraining so that feature weighting is related to the predictive sufficiency of selected montages. Retraining with selected frequency bands further characterizes spectral dependence and complementarity, allowing spatial--spectral candidate signatures to be examined in relation to their contribution to prediction.

To address these complementary modelling and interpretation objectives, this study introduces the Emotion Differential Attention Pyramid Transformer (EmoDiPyraTrans), in which structured EEG modelling, distribution regularization and spatial--spectral analysis are integrated. Sequential relative power spectral density (rPSD) graphs represent each electrode by its band-power proportions while preserving channel, frequency and temporal coordinates, enabling adaptive graph recurrence to model inter-channel relations and their temporal dynamics at different feature depths \cite{bai2020adaptive}. The resulting graph representations are refined through differential self- and cross-attention, which use differences between attention maps to enhance node features and aggregate channels while providing spatial readouts \cite{ye2025differential}. These features are subsequently combined by pyramid fusion and passed to a Differential Transformer Encoder, allowing information from different feature depths to inform temporal modelling. During model development, representation learning is regularized through maximum mean discrepancy (MMD) and confidence-filtered pseudo-label objectives, which are computed from two streams drawn from development participants.

The framework is evaluated through three complementary lines of evidence, beginning with cross-subject binary emotion decoding on SEED, FACED, MAHNOB-HCI, DEAP and DREAMER \cite{zheng2015investigating,chen2023large,soleymani2011multimodal,koelstra2011deap,katsigiannis2017dreamer}, where ablations characterize the contributions of model components and training objectives across datasets and metrics. These benchmark comparisons are extended to DEP-EEG, where four protocols assess performance within healthy and depression cohorts, under healthy-to-depression transfer and with mixed-population development. Complementing these predictive evaluations, SEED analyses combine attention readouts, trainable node masks, channel pruning and retraining with selected frequency bands to characterize model-derived candidate signatures whose predictive roles are examined through changes to the input representation. In the channel-pruning experiments, channels are selected using development participants before the resulting montages are evaluated in held-out individuals, linking spatial feature selection to cross-subject prediction. Together, these evaluations characterize performance across participants and populations while identifying spatial--spectral features that support emotion decoding on SEED.

\section{Methods}
\label{method}
\subsection{EEG preprocessing and sequential rPSD graph construction}
\label{subsec:rpsd}

Because emotion-discriminative EEG information is distributed across frequency bands and scalp channels~\cite{zheng2015investigating}, sequential rPSD graphs~\cite{ding2025emt} were constructed to preserve spectral composition, channel identity and temporal order. To represent this structure, EEG trials were divided into overlapping segments and subsegments, with dataset-specific window settings given in Section~\ref{subsubsec:datasets}. Within each subsegment, channel-wise power spectral density was estimated using Welch's method and integrated over delta (1--4~Hz), theta (4--8~Hz), alpha (8--12~Hz), low beta (12--16~Hz), beta (16--20~Hz), high beta (20--28~Hz) and gamma (30--45~Hz). The selected bands formed $\mathcal{B}=\{B_1,\ldots,B_F\}$, with $F=7$ by default and $F=6$ for DEAP, while relative spectral composition was emphasized by normalizing band power $P_{m,t,c,b}$ as $R_{m,t,c,b}=P_{m,t,c,b}/(\sum_{b'=1}^{F}P_{m,t,c,b'}+\epsilon)$, where $m$, $t$, $c$ and $b$ index segments, subsegments, channels and bands, respectively, and $\epsilon$ ensures numerical stability.

For each subsegment, the graph was defined as $\mathcal{G}_{m,t}=(\mathcal{V},G_{m,t})$, where $\mathcal{V}$ denotes the $C$ channel nodes and $G_{m,t}\in\mathbb{R}^{C\times F}$ contains their rPSD features. Ordering these graphs over $T$ subsegments yielded the input tensor $\mathcal{G}_m\in\mathbb{R}^{C\times T\times F}$, preserving channel, frequency and temporal coordinates for subsequent modelling.

\subsection{Architecture of EmoDiPyraTrans}
\label{subsec:architecture}

\begin{figure*}
      \centering
      \includegraphics[width=\textwidth]{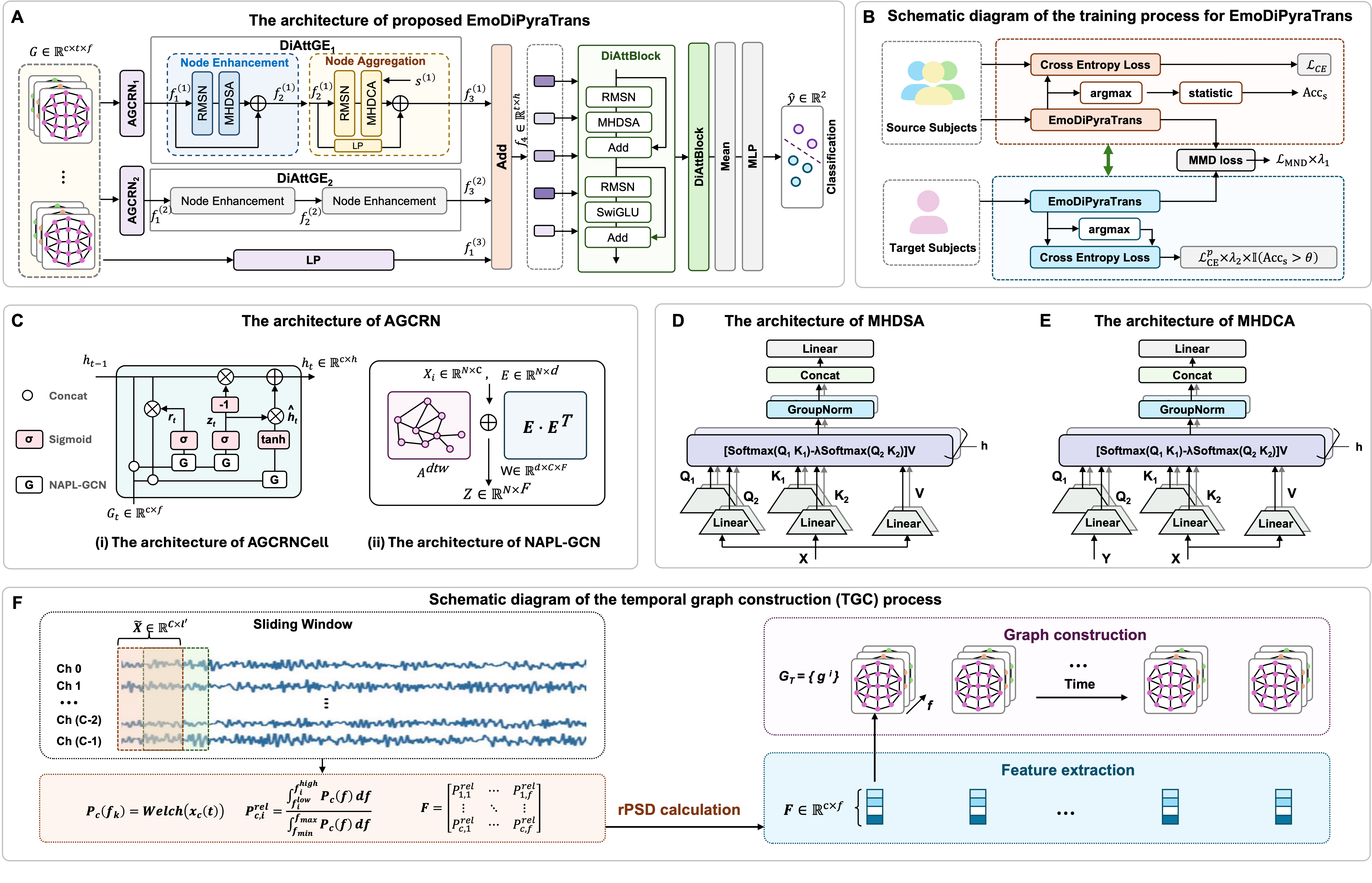}
      \caption{\textbf{Overview of EmoDiPyraTrans, its development procedure and temporal rPSD-graph construction.} \textbf{A,} Raw EEG is converted to relative power spectral density (rPSD) graphs $\mathcal{G}\in\mathbb{R}^{C\times T\times F}$ (channels $\times$ subsegments $\times$ frequency bands), which are encoded by parallel one- and two-layer adaptive graph convolutional recurrent network (AGCRN) branches and a linear projection. The graph representations are processed by Differential Attention-based Graph Enhancement modules (DiAttGE$_1$ and DiAttGE$_2$) before pyramid fusion, stacked Differential Attention Blocks and a multilayer perceptron (MLP) classifier are applied. \textbf{B,} After participants are assigned to development and outer-test sets, development data are partitioned into source-training and target-validation streams according to their roles in model development. Source labels supply cross-entropy, while target-validation signals supply maximum mean discrepancy (MMD) and confidence-filtered pseudo-label losses; target-validation labels are reserved for validation accuracy and model selection, with protocol-specific partitions detailed in Section~\ref{subsubsec:evaluation}. \textbf{C,} AGCRN, comprising an AGCRNCell and NAPL-GCN with adaptive adjacency from learnable node embeddings \cite{bai2020adaptive}. \textbf{D,} Multi-Head Differential Self-Attention (MHDSA) for feature refinement \cite{ye2025differential}. \textbf{E,} Multi-Head Differential Cross-Attention (MHDCA) for query-driven channel aggregation and its channel-attention output \cite{ye2025differential}. \textbf{F,} Temporal graph construction using sliding-window segmentation and Welch rPSD estimation.}
      \label{fig:model}
\end{figure*}

\subsubsection{Adaptive graph recurrence for spatial--temporal encoding}
Emotion processing involves distributed brain systems and temporally structured neural responses~\cite{mohammadi2023brain,bo2022decoding}, motivating joint modelling of inter-channel dependencies and evolving EEG representations. Accordingly, parallel one- and two-layer AGCRNs~\cite{bai2020adaptive} were introduced to encode sequential rPSD graphs at complementary feature depths, with each branch learning a channel-embedding matrix $E^{(r)}\in\mathbb{R}^{C\times D_e}$ for $r\in\{1,2\}$. The resulting adjacency and node-adaptive graph convolution were expressed as
\begin{equation}
\label{eq:adaptive_graph_operator}
\begin{aligned}
 A^{(r)} &= \operatorname{softmax}_{\mathrm{row}}
 \!\left[\operatorname{ReLU}\!\left(E^{(r)}E^{(r)\top}\right)\right],\\
 [\mathcal{K}_{E^{(r)}}(X)]_c
 &=\sum_{k=0}^{K-1}[\mathsf{T}_k(A^{(r)})X]_c W_{c,k}^{(r)}+b_c^{(r)},
\end{aligned}
\end{equation}
where $X$ contains channel features, $\mathsf{T}_k$ is the $k$th Chebyshev polynomial and $K$ is the filter order, with node-specific weights and biases generated linearly from channel embeddings. Within each branch, the learned adjacency provides the relational structure for gated recurrent state updates, producing $f_1^{(r)}\in\mathbb{R}^{C\times T\times H}$ as successive subsegments are processed. To carry channel--frequency information directly into fusion, a parallel linear projection maps the flattened input at each subsegment to $f_1^{(3)}\in\mathbb{R}^{T\times H}$.

\subsubsection{Differential attention for selective node enhancement}
Scalp EEG contains emotion-related activity alongside recording noise whose contribution depends on sensor conditions~\cite{kappenman2010effects}, motivating selective feature refinement. Differential self-attention was therefore introduced to emphasize discriminative context through a learned signed contrast between attention distributions~\cite{ye2025differential}. For attention head $h$, this contrast and the associated value aggregation were defined as
\begin{equation}
\label{eq:differential_attention_kernel}
\begin{aligned}
 D_h
 &=\operatorname{softmax}\!\left(
 \frac{Q_{h,1}K_{h,1}^{\top}/\sqrt{d_h}+M}{\tau}\right)\\
 &\quad-\lambda_h\operatorname{softmax}\!\left(
 \frac{Q_{h,2}K_{h,2}^{\top}/\sqrt{d_h}+M}{\tau}\right),\\
 O_h&=\operatorname{Dropout}(D_h)V_h,
\end{aligned}
\end{equation}
where $d_h=H/H_a$, $\lambda_h$ is a learned head-specific coefficient, and $M$ and $\tau$ specify the attention mask and temperature, respectively. Queries and keys were each split into two $d_h$-dimensional projections per head, while the $2d_h$-dimensional value outputs were RMS-normalized, rescaled by $1-\lambda_{\mathrm{init}}$, concatenated, projected to dimension $H$ and subjected to output dropout. With RMSNorm preceding MHDSA and a residual path preserving the incoming representation, this operation formed the node-enhancement stage of each DiAttGE branch (Fig.~\ref{fig:model}D).

\subsubsection{Query-guided aggregation of complementary channel features}
Because EEG channels contribute unequally to emotion discrimination~\cite{zheng2015investigating}, each DiAttGE branch used MHDCA with a learned query $s^{(r)}\in\mathbb{R}^{1\times H}$ shared across subsegments, yielding input-dependent channel weights and a temporal embedding (Fig.~\ref{fig:model}E). For the subsegment-wise slice $F_t^{(r)}\in\mathbb{R}^{C\times H}$ of the AGCRN output $f_1^{(r)}$, the coupled enhancement and aggregation stages were written as
\begin{equation}
\label{eq:diattge_composition}
\begin{aligned}
 U_t^{(r)}
 &=F_t^{(r)}+\mathcal{D}_{\mathrm{self}}^{(r)}
 \!\left(\mathcal{N}(F_t^{(r)})\right),\\
 z_t^{(r)}
 &=s^{(r)}+\mathcal{D}_{\mathrm{cross}}^{(r)}
 \!\left(\mathcal{N}(s^{(r)}),U_t^{(r)}\right)\\
 &\quad+\mathcal{P}_r\!\left(\operatorname{vec}(U_t^{(r)})\right),
\end{aligned}
\end{equation}
where $\mathcal{N}$ denotes independently parameterized RMSNorm layers, $\mathcal{D}_{\mathrm{self}}$ and $\mathcal{D}_{\mathrm{cross}}$ are the multi-head operators built from Eq.~\eqref{eq:differential_attention_kernel}, and $\mathcal{P}_r$ projects the flattened channel features to dimension $H$ with dropout. The projected residual provides a direct channel-feature route, while stacking $z_t^{(r)}$ over time yields $f_3^{(r)}\in\mathbb{R}^{T\times H}$ and retaining $D_h$ supplies the channel weights used for interpretation.

\subsubsection{Pyramid fusion and contextual temporal integration}
Complementary feature depths expose different compositions of channel information, while temporal context relates successive EEG representations during emotion elicitation~\cite{ding2025emt}. To integrate these contributions, pyramid fusion averaged the two graph-enhanced branches and the direct projection, after which a Differential Transformer Encoder contextualized the resulting sequence before classification:
\begin{equation}
\label{eq:pyramid_temporal_readout}
\begin{aligned}
 f_4&=\tfrac{1}{3}\left(f_3^{(1)}+f_3^{(2)}+f_1^{(3)}\right),\\
 Z&=\mathcal{N}_{\mathrm{out}}\!\left[
 (\mathcal{T}_L\circ\cdots\circ\mathcal{T}_1)(f_4)\right],\\
 p&=\operatorname{softmax}\!\left[
 \operatorname{MLP}\!\left(\frac{1}{T}\sum_{t=1}^{T}Z_t\right)\right].
\end{aligned}
\end{equation}
Within an encoder of $L$ blocks, each $\mathcal{T}_\ell$ applied pre-normalized MHDSA and a SwiGLU feed-forward transformation through successive residual updates, while $\mathcal{N}_{\mathrm{out}}$ normalized the final sequence $Z\in\mathbb{R}^{T\times H}$. Temporal mean pooling then supplied the two-layer MLP, whose output $p\in\mathbb{R}^{2}$ contains the class probabilities.

\subsection{Cross-subject distribution regularization within the development fold}
\label{subsec:regularization}

Individual differences in the physical properties shaping scalp EEG~\cite{antonakakis2020inter} motivate learning representations across participants with complementary objectives for class discrimination and distribution regularization. Accordingly, model development used source-training and target-validation streams drawn from development participants, as detailed in Section~\ref{subsubsec:evaluation}. A shared encoder and classifier processed both streams, with source labels supplying label-smoothed cross-entropy $\mathcal{L}_{\mathrm{CE}}$ and the encoded mini-batches supplying squared empirical MMD $\mathcal{L}_{\mathrm{MMD}}$, including within-batch diagonal terms, for distribution regularization.

To extend class-discriminative supervision to confidently predicted target-validation samples, pseudo-label learning was activated when source mini-batch accuracy exceeded $\tau_a$, with samples contributing when their maximum class probability exceeded $\tau_c$. Their most probable classes served as pseudo-labels, with $\mathcal{L}_{\mathrm{PL}}$ defined as mean cross-entropy over qualifying samples and set to zero when none qualified; the three objectives were combined as
\begin{equation}
\label{eq:training_objective}
    \mathcal{L}_{\mathrm{total}}
    =
    \mathcal{L}_{\mathrm{CE}}
    +
    \lambda_{\mathrm{MMD}}\mathcal{L}_{\mathrm{MMD}}
    +
    \lambda_{\mathrm{PL}}\mathcal{L}_{\mathrm{PL}}.
\end{equation}

\subsection{Interpretability analysis and model-derived candidate signatures}
\label{subsec:interpretability}

\subsubsection{Cross-attention-based spatial readouts}
\label{subsubsec:spatial_attention}

To characterize the decoder's channel weighting across emotion conditions, spatial analysis used MHDCA weights from both DiAttGE branches on SEED, whose positive--negative contrast was represented by 62 channels and seven frequency bands. For the overview map in Fig.~\ref{fig:space_biomarker}A, branch weights were averaged over participants, samples, time, heads and query seeds before the two channel vectors were multiplied and min--max scaled to $[0,1]$; repeating this procedure within participants yielded the vectors whose across-participant standard deviation is shown in panel B.

Condition-resolved scores $u_{n,i,c}$ for participant $n$, sample $i$ and channel $c$ were obtained by multiplying branch weights before averaging over time, heads and query seeds. The pooled negative-minus-positive contrast was scaled to $[-1,1]$ for panel C, whereas panel D summarized the standard deviation of separately scaled sample-level contrasts after the pooled condition lists were truncated to equal length and paired in their saved order.

\subsubsection{Trainable node masks and integrated spatial readouts}
\label{subsubsec:node_mask}

To identify compact channel sets that support classification by the trained predictor, a mask $m_{f,k}\in\mathbb{R}^{C}$ was fitted using data from each participant $k$ in the development set $\mathcal{D}_f$ of outer LOSO fold $f$, while the predictor remained fixed and sigmoid-transformed mask values multiplied the corresponding channel inputs. The mask objective combined classification cross-entropy with sparsity and entropy penalties promoting sparse, near-binary activation, together with a Kullback--Leibler term constraining mean activation towards $\rho$; participant-wise normalization then yielded the channel-selection scores $\tilde{m}_{f,k,c}$.

For visualization, fold-averaged participant masks were sigmoid-transformed and min--max scaled to $\bar{m}_{n,c}\in[0,1]$, with their across-participant mean and standard deviation shown in Fig.~\ref{fig:space_biomarker}E,F. To combine masks with condition-specific attention, each participant's negative and positive sequences were truncated to their common length $J_n$ while retaining saved order, yielding the integrated score
\begin{equation}
\label{eq:integrated_spatial_score}
    B_{n,c}
    =
    \frac{\bar{m}_{n,c}}{J_n}
    \sum_{j=1}^{J_n}
    \left(u_{n,j,c}^{-}-u_{n,j,c}^{+}\right).
\end{equation}
Each participant's channel vector was then min--max scaled to $[-1,1]$, giving $\widetilde{B}_{n,c}$ with zero at the score-range midpoint. Its across-participant mean and standard deviation are shown in panels G,H, with population standard deviations used throughout the displayed dispersion maps.

\subsubsection{Statistical assessment and channel pruning}
\label{subsubsec:channel_pruning}

Across 15 participants, two-sided one-sample $t$-tests compared $\widetilde{B}_{n,c}$ with zero, followed by Benjamini--Hochberg correction over 62 channels for Fig.~\ref{fig:space_biomarker}J; panel I shows the mean scaled score divided by its sample standard deviation. Exploratory two-sided paired $t$-tests compared raw fused-attention means computed from each participant's full condition-specific sample sets, with Benjamini--Hochberg correction across the ten channels having the smallest corrected $q$ values in panel J, selected and tested within the same cohort (panel L).

Predictive sufficiency was assessed through fold-specific channel pruning, with threshold $\theta\in[0,1]$ retaining channels that met the criterion in every development participant:
\begin{equation}
\label{eq:pruned_channels}
    \mathcal{C}_{\theta}^{(f)}
    =
    \left\{
    c \mid \tilde{m}_{f,k,c}\geq \theta,\ \forall k\in\mathcal{D}_f
    \right\}.
\end{equation}
Mask fitting, threshold determination and channel selection used development participants, after which the model was retrained on selected channels and evaluated on the held-out participant under SEED LOSO. Because selections varied between folds, retained-channel counts were averaged across folds (Fig.~\ref{fig:space_biomarker}K,M).

\subsubsection{Frequency-band contribution and spectral analysis}
\label{subsubsec:spectral_analysis}

To connect band-specific EEG representations with their predictive contribution, spectral dependence was assessed on SEED by retraining leave-one-band-out, single-band, two-band and reduced multiband models under the same participant-level folds. Removal effects were measured relative to the full seven-band model, while selected-band configurations characterized standalone discrimination and spectral complementarity.

For each spectral configuration, branch-specific MHDCA weights were first averaged over time, heads and query seeds, after which the two branch vectors were multiplied. Positive-minus-negative contrasts were then computed from each participant's condition means and averaged over the 15 participants, yielding configuration-specific spatial maps. Map similarity was quantified by Pearson correlation, while regional mean absolute contrasts summarized frontal, frontocentral, central, centroparietal, parietal, parieto-occipital and occipital weighting.

\subsection{Experimental setup}
\label{subsec:experimental_setup}
\subsubsection{Datasets and preprocessing}
\label{subsubsec:datasets}

EmoDiPyraTrans was evaluated on five public EEG emotion benchmarks and the clinical DEP-EEG dataset, with binary emotion classification assessed on held-out participants throughout. SEED~\cite{zheng2015investigating} comprises 15 participants and 15 Chinese film clips eliciting positive, negative and neutral emotions; analysis used positive and negative trials from 62 channels processed at 200~Hz.

FACED~\cite{chen2023large} comprises 123 participants viewing 28 emotion-inducing videos, with 32 channels sampled at 250~Hz; released EEG and annotation files supplied the signals and valence scores, respectively. Scores above 3.0 defined high valence, while scores at or below 3.0 defined low valence.

MAHNOB-HCI~\cite{soleymani2011multimodal} included 27 participants with 32 EEG channels at 128~Hz; signals were average-referenced, cropped by 30~s at each end, band-pass filtered at 0.3--45~Hz and notch-filtered at 50~Hz. SAM-like valence ratings on the 1--9 scale defined high and low valence above and below 5, respectively, with threshold-valued samples discarded.

DEAP~\cite{koelstra2011deap} used official preprocessed recordings from 32 participants with 32 channels at 128~Hz, removing each trial's 3-s baseline and assigning scores above 5.0 to high valence and the remainder to low valence. Because the released signals were filtered at 4--45~Hz, rPSD used the six bands above $\delta$ defined in Section~\ref{subsec:rpsd}.

DREAMER~\cite{katsigiannis2017dreamer} comprised 23 participants completing 18 audiovisual trials each, with analysis retaining the final 60~s from 14 channels at 128~Hz; scores above 3.0 defined high valence and the remainder low valence.

DEP-EEG~\cite{huang2021neurofeedback,guan2025method} comprised 40 healthy controls and 20 participants with depression, each viewing four neutral and four positive videos in randomised category order. Analysis retained 50~s of approximately 60~s released per video from 30 channels recorded at 250~Hz with the right mastoid A2 online reference; released signals had undergone common-average re-referencing, 0.01-Hz high-pass filtering and power-line-noise suppression.

All trials were converted to sequential rPSD graphs using 20-s windows with 80\% overlap and a 4-s hop, as described in Section~\ref{subsec:rpsd}; subsegments overlapped by 75\% and lasted 2~s for SEED and DREAMER or 4~s for the remaining datasets.

\subsubsection{Evaluation protocol and baseline methods}
\label{subsubsec:evaluation}

Outer-test participants were separated before development data were partitioned into the source-training and target-validation streams defined in Section~\ref{subsec:regularization}. SEED, MAHNOB-HCI, DEAP and DREAMER used leave-one-subject-out (LOSO) cross-validation, whereas FACED used ten outer folds with 12 test participants in the first nine and 15 in the last. Target-validation labels were reserved for computing validation accuracy and selecting the model, with the parameter set achieving the highest validation accuracy evaluated on the corresponding outer-test fold.

For public datasets, HC-LOSO, DEP-LOSO and HC$\rightarrow$DEP, each development participant underwent a class-stratified 80:20 source-training/target-validation split, except FACED, which used 90:10; random seed 2022 fixed the resulting sample-index assignments across models and training runs.

DEP-EEG positive-versus-neutral decoding used HC-LOSO and DEP-LOSO to hold out one of 40 healthy controls or 20 participants with depression, respectively, with both development streams drawn from the remaining participants. For cross-population transfer, HC$\rightarrow$DEP developed one model on all healthy controls and evaluated it separately on each participant with depression. Mixed-population DEP-LOSO (HC+DEP$\rightarrow$DEP LOSO in Table~\ref{tab:dep_eeg_comparison}) comprised 20 outer folds, each holding out one participant with depression and all 64 corresponding segments. The remaining 3,776 segments from 40 controls and 19 participants with depression were pooled and jointly stratified by population and emotion class into training ($n=3,020$) and validation ($n=756$) subsets using an 8:2 segment-level split and seed 2022.

Accuracy (Acc) and positive-class F1 were calculated per test participant from sample-level predictions, with class 1 denoting positive emotion or high valence and class 0 the complementary emotion class; F1 was zero when positive predictions were absent. Reported means were unweighted participant averages, while standard deviations described between-participant variation, using divisor $n=20$ for mixed-population DEP-LOSO. Comparators included recurrent, convolutional, graph-based, distribution-learning, Transformer and recent spatiotemporal EEG architectures, whose citations are listed in Tables~\ref{tab:public_benchmarks} and \ref{tab:dep_eeg_comparison}, with rankings based on participant-mean results under the corresponding protocols.

\subsubsection{Implementation details and hyperparameter settings}
\label{subsubsec:implementation}

Experiments used PyTorch 2.1.1, CUDA 11.8 and Python 3.10.12 on Ubuntu 22.04 with one NVIDIA GeForce RTX 4080 GPU, while the random seed was fixed at 2022. Models were trained using AdamW with mini-batches of 128, learning rate $3\times10^{-4}$, dropout 0.25 and label smoothing 0.1; training continued for at most 100 epochs on SEED and 50 on the other datasets, with early-stopping patience of ten epochs.

The model used hidden graph dimension $H=32$, AGCRN node-embedding dimension 10, graph-convolution order $K=4$ and $H_a=8$ attention heads, with eight Differential Transformer Encoder layers for all datasets except DREAMER, which used four. The primary SEED configuration was fixed before evaluation, with subsequent scans assessing hyperparameter sensitivity independently of the primary model's architecture and parameter selection. DiAttGE used differential-coefficient offset $\lambda_{\mathrm{init}}=0.2$, with the learned head-wise coefficient parameterized as $\lambda_h=\exp(a_{h,1}^{\top}b_{h,1})-\exp(a_{h,2}^{\top}b_{h,2})+\lambda_{\mathrm{init}}$ for trainable vectors $a$ and $b$. The encoder offset followed $\lambda_{\mathrm{init}}^{(l)}=0.8-0.6\exp[-0.3(l-1)]$, while attention temperature was 1.0 for MHDSA and 0.5 for MHDCA and RMSNorm used $\epsilon=10^{-5}$. MHDSA applied a lower-triangular mask over the attended axis, corresponding to channel order in DiAttGE and subsegment order in the temporal encoder, while MHDCA attended to all channel keys. The two-layer MLP classifier projected 32 to 16 features with ReLU and batch normalization, followed by a 16-to-2 output layer.

MMD and pseudo-label losses each had weight 0.5, with the MMD radial basis function kernel combining five Gaussian kernels, a multiplier of 2.0 and adaptive bandwidth from pairwise feature distances; pseudo-label source-accuracy and target-confidence thresholds were $\tau_a=0.8$ and $\tau_c=0.9$. SEED node-mask sparsity, entropy and Kullback--Leibler-divergence weights were 5.0, 0.1 and 1.0, respectively, with sparse target activation rate $\rho=0.05$.

\section{Results}

\subsection{Cross-subject decoding across public datasets}

Across the five public datasets, EmoDiPyraTrans achieved the highest participant-mean accuracy and positive-class F1 among the evaluated methods (Table~\ref{tab:public_benchmarks}), with the magnitude of these gains varying across datasets. On SEED, accuracy and F1 reached $0.928\pm0.055$ and $0.928\pm0.054$, respectively, corresponding to gains of 9.4 and 7.8 percentage points over MSGM and lower between-participant standard deviations. On MAHNOB-HCI, accuracy and F1 reached $0.714\pm0.079$ and $0.611\pm0.154$, respectively, exceeding the strongest corresponding baseline values by 12.7 and 21.5 points.

For FACED and DEAP, F1 gains were 0.8 and 0.3 percentage points, respectively, whereas the corresponding accuracy gains over the strongest comparators were 1.1 and 7.0 points. On DREAMER, accuracy reached $0.671\pm0.066$ and positive-class F1 reached $0.328\pm0.273$, exceeding the strongest corresponding baselines by 5.7 and 6.5 percentage points, respectively.

\begin{table*}[t]
\centering
\caption{Cross-subject emotion classification across five public EEG datasets.}
\label{tab:public_benchmarks}
\begingroup
\footnotesize\setlength{\tabcolsep}{2pt}
\renewcommand{\arraystretch}{1.05}
\begin{tabular*}{\textwidth}{@{\extracolsep{\fill}}lcccccccccc}
\toprule
\multirow{2}{*}{Method} & \multicolumn{2}{c}{SEED} & \multicolumn{2}{c}{FACED} & \multicolumn{2}{c}{MAHNOB-HCI} & \multicolumn{2}{c}{DEAP} & \multicolumn{2}{c}{DREAMER}\\
\cmidrule(lr){2-3}\cmidrule(lr){4-5}\cmidrule(lr){6-7}\cmidrule(lr){8-9}\cmidrule(lr){10-11}
& Acc & F1 & Acc & F1 & Acc & F1 & Acc & F1 & Acc & F1 \\
\midrule
LSTM~\cite{soleymani2015analysis} & \shortstack{$0.733$\\$\pm0.158$} & \shortstack{$0.670$\\$\pm0.274$} & \shortstack{$0.568$\\$\pm0.063$} & \shortstack{$0.700$\\$\pm0.064$} & \shortstack{$0.585$\\$\pm0.126$} & \shortstack{$0.341$\\$\pm0.274$} & \shortstack{$0.517$\\$\pm0.099$} & \shortstack{$0.570$\\$\pm0.219$} & \shortstack{$0.614$\\$\pm0.106$} & \shortstack{$0.125$\\$\pm0.193$} \\
BiDANN~\cite{li2018bi} & \shortstack{$0.794$\\$\pm0.165$} & \shortstack{$0.777$\\$\pm0.159$} & \shortstack{$0.634$\\$\pm0.070$} & \shortstack{$0.738$\\$\pm0.064$} & --- & --- & --- & --- & --- & --- \\
DGCNN~\cite{song2018eeg} & \shortstack{$0.724$\\$\pm0.145$} & \shortstack{$0.619$\\$\pm0.311$} & \shortstack{$0.562$\\$\pm0.045$} & \shortstack{$0.697$\\$\pm0.043$} & \shortstack{$0.584$\\$\pm0.104$} & \shortstack{$0.376$\\$\pm0.267$} & \shortstack{$0.501$\\$\pm0.094$} & \shortstack{$0.495$\\$\pm0.275$} & \shortstack{$0.577$\\$\pm0.116$} & \shortstack{$0.169$\\$\pm0.189$} \\
GCB-Net~\cite{zhang2019gcb} & \shortstack{$0.684$\\$\pm0.172$} & \shortstack{$0.517$\\$\pm0.357$} & \shortstack{$0.565$\\$\pm0.052$} & \shortstack{$0.685$\\$\pm0.053$} & --- & --- & --- & --- & --- & --- \\
RGNN~\cite{zhong2020eeg} & \shortstack{$0.790$\\$\pm0.148$} & \shortstack{$0.802$\\$\pm0.133$} & \shortstack{$0.587$\\$\pm0.050$} & $0.722^{\dagger}$ & --- & --- & --- & --- & --- & --- \\
DMATN~\cite{wang2021deep} & \shortstack{$0.773$\\$\pm0.155$} & \shortstack{$0.772$\\$\pm0.163$} & \shortstack{$0.614$\\$\pm0.049$} & \shortstack{$0.683$\\$\pm0.069$} & --- & --- & --- & --- & --- & --- \\
TSception~\cite{ding2022tsception} & \shortstack{$0.662$\\$\pm0.181$} & \shortstack{$0.621$\\$\pm0.283$} & \shortstack{$0.619$\\$\pm0.088$} & \shortstack{$0.702$\\$\pm0.237$} & \shortstack{$0.580$\\$\pm0.108$} & \shortstack{$0.338$\\$\pm0.219$} & \shortstack{$0.546$\\$\pm0.067$} & \shortstack{$0.610$\\$\pm0.226$} & \shortstack{$0.597$\\$\pm0.102$} & \shortstack{$0.198$\\$\pm0.163$} \\
TCN~\cite{zhang2022visual} & \shortstack{$0.765$\\$\pm0.140$} & \shortstack{$0.737$\\$\pm0.219$} & \shortstack{$0.552$\\$\pm0.035$} & \shortstack{$0.673$\\$\pm0.035$} & \shortstack{$0.585$\\$\pm0.128$} & \shortstack{$0.344$\\$\pm0.260$} & \shortstack{$0.547$\\$\pm0.085$} & \shortstack{$0.564$\\$\pm0.212$} & \shortstack{$0.599$\\$\pm0.094$} & \shortstack{$0.263$\\$\pm0.198$} \\
Conformer~\cite{song2022eeg} & \shortstack{$0.612$\\$\pm0.127$} & \shortstack{$0.529$\\$\pm0.221$} & \shortstack{$0.590$\\$\pm0.035$} & \shortstack{$0.720$\\$\pm0.035$} & \shortstack{$0.530$\\$\pm0.087$} & \shortstack{$0.140$\\$\pm0.159$} & \shortstack{$0.542$\\$\pm0.094$} & \shortstack{$0.676$\\$\pm0.144$} & \shortstack{$0.579$\\$\pm0.106$} & \shortstack{$0.147$\\$\pm0.138$} \\
AMDET~\cite{xu2023amdet} & \shortstack{$0.721$\\$\pm0.168$} & \shortstack{$0.649$\\$\pm0.306$} & \shortstack{$0.591$\\$\pm0.043$} & \shortstack{$0.726$\\$\pm0.043$} & --- & --- & --- & --- & --- & --- \\
PGCN~\cite{jin2024pgcn} & \shortstack{$0.759$\\$\pm0.184$} & \shortstack{$0.741$\\$\pm0.190$} & \shortstack{$0.558$\\$\pm0.078$} & \shortstack{$0.665$\\$\pm0.085$} & --- & --- & --- & --- & --- & --- \\
EmT~\cite{ding2025emt} & \shortstack{$0.802$\\$\pm0.115$} & \shortstack{$0.821$\\$\pm0.093$} & \shortstack{$0.608$\\$\pm0.065$} & \shortstack{$0.740$\\$\pm0.058$} & \shortstack{$0.587$\\$\pm0.125$} & \shortstack{$0.396$\\$\pm0.248$} & \shortstack{$0.514$\\$\pm0.086$} & \shortstack{$0.602$\\$\pm0.183$} & \shortstack{$0.599$\\$\pm0.107$} & \shortstack{$0.192$\\$\pm0.202$} \\
MSGM~\cite{liu2026msgm} & \shortstack{$0.834$\\$\pm0.114$} & \shortstack{$0.850$\\$\pm0.091$} & \shortstack{$0.632$\\$\pm0.036$} & \shortstack{$0.760$\\$\pm0.037$} & --- & --- & --- & --- & --- & --- \\
\midrule
EmoDiPyraTrans & \shortstack{$\mathbf{0.928}$\\$\mathbf{\pm0.055}$} & \shortstack{$\mathbf{0.928}$\\$\mathbf{\pm0.054}$} & \shortstack{$\mathbf{0.645}$\\$\mathbf{\pm0.048}$} & \shortstack{$\mathbf{0.768}$\\$\mathbf{\pm0.050}$} & \shortstack{$\mathbf{0.714}$\\$\mathbf{\pm0.079}$} & \shortstack{$\mathbf{0.611}$\\$\mathbf{\pm0.154}$} & \shortstack{$\mathbf{0.617}$\\$\mathbf{\pm0.073}$} & \shortstack{$\mathbf{0.679}$\\$\mathbf{\pm0.110}$} & \shortstack{$\mathbf{0.671}$\\$\mathbf{\pm0.066}$} & \shortstack{$\mathbf{0.328}$\\$\mathbf{\pm0.273}$} \\
\bottomrule
\end{tabular*}
\endgroup
\par\vspace{2pt}
\begin{minipage}{\textwidth}
\footnotesize\raggedright
\textbf{Note.} SEED and FACED baselines were taken from published comparisons in EmT~\cite{ding2025emt} and MSGM~\cite{liu2026msgm}, while MAHNOB-HCI, DEAP and DREAMER baselines were rerun under the corresponding protocols. Each cell shows the unweighted participant mean above its $\pm$ s.d., with bold denoting the highest value per column and an em dash indicating an unavailable value. Acc denotes accuracy, while F1 denotes positive-class F1 (positive emotion for SEED; high valence otherwise), which was set to 0 when the positive prediction count was zero. $\dagger$The RGNN FACED F1 entry reports the source mean, with the reported s.d. of 0.721 flagged as inconsistent with the bounded metric.
\end{minipage}
\end{table*}

\subsection{Emotion decoding within and across clinical populations}

\begin{table*}[t]
\centering
\caption{Cross-subject positive-versus-neutral classification on DEP-EEG.}
\label{tab:dep_eeg_comparison}
\resizebox{0.94\textwidth}{!}{%
\begin{tabular}{lcccccccc}
\toprule
\multirow{2}{*}{\textbf{Method}} & \multicolumn{2}{c}{\textbf{HC-LOSO}} & \multicolumn{2}{c}{\textbf{DEP-LOSO}} & \multicolumn{2}{c}{\textbf{HC$\rightarrow$DEP}} & \multicolumn{2}{c}{\textbf{HC+DEP$\rightarrow$DEP LOSO}} \\
\cmidrule(lr){2-3} \cmidrule(lr){4-5} \cmidrule(lr){6-7} \cmidrule(lr){8-9}
& \textbf{Acc} & \textbf{F1} & \textbf{Acc} & \textbf{F1} & \textbf{Acc} & \textbf{F1} & \textbf{Acc} & \textbf{F1} \\
\midrule
LSTM~\cite{soleymani2015analysis} & $0.650\pm0.124$ & $0.562\pm0.264$ & $0.615\pm0.114$ & $0.576\pm0.243$ & $0.557$ & $0.430$ & $0.537\pm0.092$ & $0.345\pm0.280$ \\
DGCNN~\cite{song2018eeg} & $0.687\pm0.137$ & $0.644\pm0.248$ & $0.577\pm0.121$ & $0.513\pm0.263$ & $0.528$ & $0.385$ & $0.532\pm0.059$ & $0.312\pm0.260$ \\
TSception~\cite{ding2022tsception} & $0.647\pm0.125$ & $0.589\pm0.241$ & $0.548\pm0.094$ & $0.455\pm0.211$ & $0.516$ & $0.389$ & $0.548\pm0.069$ & $0.409\pm0.232$ \\
TCN~\cite{zhang2022visual} & $0.636\pm0.127$ & $0.561\pm0.254$ & $0.652\pm0.137$ & $0.607\pm0.269$ & $0.570$ & $0.460$ & $\mathbf{0.600\pm0.100}$ & $0.476\pm0.261$ \\
Conformer~\cite{song2022eeg} & $0.610\pm0.100$ & $0.481\pm0.249$ & $0.532\pm0.089$ & $0.568\pm0.119$ & $0.516$ & $0.148$ & $0.574\pm0.059$ & $0.429\pm0.192$ \\
EmT~\cite{ding2025emt} & $0.655\pm0.122$ & $0.625\pm0.209$ & $0.615\pm0.145$ & $0.566\pm0.255$ & $0.569$ & $0.416$ & $0.586\pm0.116$ & $0.482\pm0.223$ \\
\midrule
EmoDiPyraTrans & $\mathbf{0.802\pm0.130}$ & $\mathbf{0.797\pm0.171}$ & $\mathbf{0.704\pm0.094}$ & $\mathbf{0.709\pm0.109}$ & $\mathbf{0.591}$ & $\mathbf{0.598}$ & $0.581\pm0.114$ & $\mathbf{0.498\pm0.211}$ \\
\bottomrule
\end{tabular}%
}
\par\vspace{2pt}
\begin{minipage}{0.94\textwidth}
\footnotesize\raggedright
\textbf{Note.} HC, healthy control; DEP, participant with depression; LOSO, leave-one-subject-out; Acc, accuracy. Metrics were computed per participant and are reported as mean $\pm$ s.d., while HC$\rightarrow$DEP shows the unweighted mean of 20 DEP participants. F1 denotes positive-emotion F1, which equals 0 when the positive prediction count is zero; bold denotes the highest value per protocol--metric column.
\end{minipage}
\end{table*}

Positive-versus-neutral results on DEP-EEG were compared across HC-LOSO, DEP-LOSO, HC$\rightarrow$DEP and mixed-population DEP-LOSO (Table~\ref{tab:dep_eeg_comparison}), with EmoDiPyraTrans achieving the highest accuracy and F1 in both within-population protocols. Under HC-LOSO, accuracy and F1 reached $0.802\pm0.130$ and $0.797\pm0.171$, whereas the corresponding DEP-LOSO values were $0.704\pm0.094$ and $0.709\pm0.109$. Relative to the strongest comparators, accuracy/F1 gains reached 11.5/15.3 percentage points under HC-LOSO, while DEP-LOSO yielded gains of 5.2/10.2 points.

Under HC$\rightarrow$DEP transfer, participant-mean accuracy and F1 reached 0.591 and 0.598, respectively, which placed EmoDiPyraTrans first among the evaluated methods for both metrics. Relative to HC-LOSO, these transfer scores represented decreases of 21.1 and 19.9 percentage points in accuracy and F1, respectively, while the corresponding decreases relative to DEP-LOSO were 11.3 and 11.1 points.

Mixed-population DEP-LOSO yielded $0.581\pm0.114$ accuracy and the highest positive-class F1 ($0.498\pm0.211$), whereas TCN achieved the highest accuracy at $0.600\pm0.100$. Under this mixed-population protocol, accuracy and F1 were 12.3 and 21.1 percentage points lower than under DEP-LOSO and 1.0 and 10.0 points lower than under HC$\rightarrow$DEP, with one held-out participant recording an F1 of 0.

\subsection{Model components and training-scheme ablations}

Ablations across the five public datasets and the HC-LOSO and DEP-LOSO protocols identified differential attention (DA) as the only component whose removal reduced both metrics in all seven protocols (Table~\ref{tab:ablation}), with decreases of 1.7--13.4 percentage points in accuracy and 0.4--31.0 points in F1. On SEED, replacing DA with conventional multi-head attention reduced accuracy from $0.928\pm0.055$ to $0.842\pm0.129$, while F1 decreased from $0.928\pm0.054$ to $0.830\pm0.206$ under the same substitution.

Removing MMD reduced accuracy in every protocol, with decreases ranging from 0.6 percentage points on FACED to 10.9 points on SEED, while F1 decreased on SEED, MAHNOB-HCI and both DEP-EEG LOSO protocols. On DEAP, F1 increased from 0.679 to 0.686 following MMD removal, whereas the corresponding values on FACED and DREAMER remained unchanged to three decimal places.

Node enhancement (NE) and node aggregation (NA) showed dataset-dependent effects, with their individual removal reducing SEED accuracy to 0.891 and 0.921, respectively, whereas removing NA increased DEP-LOSO F1 from 0.709 to 0.737. Removing NE increased DREAMER accuracy from 0.671 to 0.685, while removing NA, either alone or with NE, increased HC-LOSO accuracy from 0.802 to 0.811.

Removing pseudo-label learning (PL) reduced SEED accuracy and F1 from 0.928 to 0.925, whereas F1 increased in five protocols, including DREAMER (0.328 to 0.377), FACED (0.768 to 0.778) and MAHNOB-HCI (0.611 to 0.616). Within SEED, removing MMD alone yielded an accuracy of 0.819, whereas the joint removal of MMD and PL yielded an accuracy of 0.844.

\begin{table*}[t]
\centering
\caption{Model components and training-scheme ablations of EmoDiPyraTrans.}
\label{tab:ablation}

\begingroup
\footnotesize\setlength{\tabcolsep}{2pt}
\renewcommand{\arraystretch}{1.05}
\begin{tabular*}{\textwidth}{@{\extracolsep{\fill}}llcccccccc}
\toprule

\multirow{2}{*}{\textbf{Dataset / Protocol}}
& \multirow{2}{*}{\textbf{Metric}}
& \multirow{2}{*}{\textbf{Full}}
& \multicolumn{4}{c}{\textbf{Component Ablation}}
& \multicolumn{3}{c}{\textbf{Training-Scheme Ablation}} \\

\cmidrule(lr){4-7}
\cmidrule(lr){8-10}

& &
& \textbf{w/o DA}\textsuperscript{a}
& \textbf{w/o NE}
& \textbf{w/o NA}
& \textbf{w/o NE \& NA}
& \textbf{w/o MMD}
& \textbf{w/o PL}
& \textbf{w/o MMD \& PL} \\

\midrule

% =========================================================
% SEED
% =========================================================
\multirow{2}{*}{SEED}
& Acc
& \shortstack{$\mathbf{0.928}$\\$\mathbf{\pm0.055}$}
& \shortstack{$0.842$\\$\pm0.129$}
& \shortstack{$0.891$\\$\pm0.100$}
& \shortstack{$0.921$\\$\pm0.058$}
& \shortstack{$0.907$\\$\pm0.078$}
& \shortstack{$0.819$\\$\pm0.123$}
& \shortstack{$0.925$\\$\pm0.051$}
& \shortstack{$0.844$\\$\pm0.106$} \\

& F1
& \shortstack{$\mathbf{0.928}$\\$\mathbf{\pm0.054}$}
& \shortstack{$0.830$\\$\pm0.206$}
& \shortstack{$0.894$\\$\pm0.096$}
& \shortstack{$0.922$\\$\pm0.058$}
& \shortstack{$0.910$\\$\pm0.072$}
& \shortstack{$0.827$\\$\pm0.117$}
& \shortstack{$0.925$\\$\pm0.051$}
& \shortstack{$0.830$\\$\pm0.139$} \\

\midrule

% =========================================================
% DEP-EEG: HC-LOSO
% =========================================================
\multirow{2}{*}{DEP-EEG (HC-LOSO)}
& Acc
& \shortstack{$0.802$\\$\pm0.130$}
& \shortstack{$0.668$\\$\pm0.149$}
& \shortstack{$0.810$\\$\pm0.131$}
& \shortstack{$\mathbf{0.811}$\\$\mathbf{\pm0.120}$}
& \shortstack{$\mathbf{0.811}$\\$\mathbf{\pm0.116}$}
& \shortstack{$0.769$\\$\pm0.134$}
& \shortstack{$0.795$\\$\pm0.134$}
& \shortstack{$0.768$\\$\pm0.130$} \\

& F1
& \shortstack{$0.797$\\$\pm0.171$}
& \shortstack{$0.583$\\$\pm0.279$}
& \shortstack{$0.788$\\$\pm0.185$}
& \shortstack{$0.804$\\$\pm0.129$}
& \shortstack{$\mathbf{0.809}$\\$\mathbf{\pm0.119}$}
& \shortstack{$0.766$\\$\pm0.157$}
& \shortstack{$0.795$\\$\pm0.169$}
& \shortstack{$0.761$\\$\pm0.163$} \\

% =========================================================
% DEP-EEG: DEP-LOSO
% =========================================================
\multirow{2}{*}{DEP-EEG (DEP-LOSO)}
& Acc
& \shortstack{$0.704$\\$\pm0.094$}
& \shortstack{$0.575$\\$\pm0.127$}
& \shortstack{$0.674$\\$\pm0.092$}
& \shortstack{$\mathbf{0.709}$\\$\mathbf{\pm0.104}$}
& \shortstack{$0.691$\\$\pm0.104$}
& \shortstack{$0.623$\\$\pm0.080$}
& \shortstack{$0.705$\\$\pm0.093$}
& \shortstack{$0.687$\\$\pm0.093$} \\

& F1
& \shortstack{$0.709$\\$\pm0.109$}
& \shortstack{$0.537$\\$\pm0.228$}
& \shortstack{$0.665$\\$\pm0.142$}
& \shortstack{$\mathbf{0.737}$\\$\mathbf{\pm0.086}$}
& \shortstack{$0.698$\\$\pm0.143$}
& \shortstack{$0.659$\\$\pm0.117$}
& \shortstack{$0.713$\\$\pm0.109$}
& \shortstack{$0.707$\\$\pm0.097$} \\

\midrule

% =========================================================
% DEAP
% =========================================================
\multirow{2}{*}{DEAP}
& Acc
& \shortstack{$\mathbf{0.617}$\\$\mathbf{\pm0.073}$}
& \shortstack{$0.597$\\$\pm0.075$}
& \shortstack{$0.603$\\$\pm0.071$}
& \shortstack{$0.613$\\$\pm0.072$}
& \shortstack{$0.614$\\$\pm0.069$}
& \shortstack{$0.609$\\$\pm0.067$}
& \shortstack{$\mathbf{0.617}$\\$\mathbf{\pm0.072}$}
& \shortstack{$0.615$\\$\pm0.074$} \\

& F1
& \shortstack{$0.679$\\$\pm0.110$}
& \shortstack{$0.588$\\$\pm0.277$}
& \shortstack{$0.655$\\$\pm0.161$}
& \shortstack{$0.666$\\$\pm0.135$}
& \shortstack{$0.644$\\$\pm0.177$}
& \shortstack{$\mathbf{0.686}$\\$\mathbf{\pm0.106}$}
& \shortstack{$0.681$\\$\pm0.114$}
& \shortstack{$0.685$\\$\pm0.110$} \\

\midrule

% =========================================================
% DREAMER
% =========================================================
\multirow{2}{*}{DREAMER}
& Acc
& \shortstack{$0.671$\\$\pm0.066$}
& \shortstack{$0.607$\\$\pm0.098$}
& \shortstack{$\mathbf{0.685}$\\$\mathbf{\pm0.068}$}
& \shortstack{$0.668$\\$\pm0.072$}
& \shortstack{$0.674$\\$\pm0.073$}
& \shortstack{$0.652$\\$\pm0.083$}
& \shortstack{$0.674$\\$\pm0.068$}
& \shortstack{$0.669$\\$\pm0.072$} \\

& F1
& \shortstack{$0.328$\\$\pm0.273$}
& \shortstack{$0.055$\\$\pm0.148$}
& \shortstack{$0.367$\\$\pm0.270$}
& \shortstack{$0.343$\\$\pm0.264$}
& \shortstack{$0.348$\\$\pm0.247$}
& \shortstack{$0.328$\\$\pm0.229$}
& \shortstack{$\mathbf{0.377}$\\$\mathbf{\pm0.259}$}
& \shortstack{$0.285$\\$\pm0.277$} \\

\midrule

% =========================================================
% FACED
% =========================================================
\multirow{2}{*}{FACED}
& Acc
& \shortstack{$\mathbf{0.645}$\\$\mathbf{\pm0.048}$}
& \shortstack{$0.628$\\$\pm0.055$}
& \shortstack{$0.639$\\$\pm0.053$}
& \shortstack{$0.642$\\$\pm0.051$}
& \shortstack{$0.642$\\$\pm0.052$}
& \shortstack{$0.639$\\$\pm0.051$}
& \shortstack{$0.642$\\$\pm0.051$}
& \shortstack{$0.638$\\$\pm0.054$} \\

& F1
& \shortstack{$0.768$\\$\pm0.050$}
& \shortstack{$0.764$\\$\pm0.048$}
& \shortstack{$0.776$\\$\pm0.041$}
& \shortstack{$0.775$\\$\pm0.042$}
& \shortstack{$0.777$\\$\pm0.040$}
& \shortstack{$0.768$\\$\pm0.051$}
& \shortstack{$\mathbf{0.778}$\\$\mathbf{\pm0.038}$}
& \shortstack{$0.770$\\$\pm0.045$} \\

\midrule

% =========================================================
% MAHNOB-HCI
% =========================================================
\multirow{2}{*}{MAHNOB-HCI}
& Acc
& \shortstack{$\mathbf{0.714}$\\$\mathbf{\pm0.079}$}
& \shortstack{$0.614$\\$\pm0.091$}
& \shortstack{$0.693$\\$\pm0.093$}
& \shortstack{$0.710$\\$\pm0.096$}
& \shortstack{$0.709$\\$\pm0.080$}
& \shortstack{$0.651$\\$\pm0.129$}
& \shortstack{$0.713$\\$\pm0.085$}
& \shortstack{$0.666$\\$\pm0.106$} \\

& F1
& \shortstack{$0.611$\\$\pm0.154$}
& \shortstack{$0.301$\\$\pm0.294$}
& \shortstack{$0.578$\\$\pm0.197$}
& \shortstack{$0.590$\\$\pm0.205$}
& \shortstack{$0.606$\\$\pm0.169$}
& \shortstack{$0.466$\\$\pm0.263$}
& \shortstack{$\mathbf{0.616}$\\$\mathbf{\pm0.159}$}
& \shortstack{$0.535$\\$\pm0.220$} \\

\bottomrule
\end{tabular*}
\endgroup
\par\vspace{2pt}
\begin{minipage}{\textwidth}
\footnotesize\raggedright
\textbf{Note.} Acc, accuracy; w/o, without; Full, complete model; DA, differential attention; NE, node enhancement; NA, node aggregation. MMD, maximum mean discrepancy-based distribution alignment; PL, pseudo-label learning; HC, healthy control; DEP, participant with depression; LOSO, leave-one-subject-out. DEP-EEG uses within-population HC-LOSO and DEP-LOSO, with each cell showing the unweighted participant mean above its $\pm$ s.d. F1 denotes positive-emotion F1 for SEED and DEP-EEG and high-valence F1 otherwise, taking a value of 0 when the positive prediction count is zero. Bold denotes the largest value per dataset--metric row, with all tied maxima shown in bold.

\par\textsuperscript{a}\,For w/o DA, differential self- and cross-attention in DiAttGE$_1$ and DiAttGE$_2$ were replaced by standard multi-head mechanisms, which also replaced differential self-attention in the DiAttBlocks.
\end{minipage}

\end{table*}

\subsection{Hyperparameter sensitivity on SEED}
SEED sensitivity analyses varied the hidden dimension, Transformer depth and graph-convolution order while holding other settings fixed, with hidden dimensions of 8, 16 and 32 yielding accuracies of $0.841\pm0.143$, $0.901\pm0.080$ and $0.928\pm0.055$, respectively. Increasing the dimension to 64 retained a mean accuracy of $0.928$ (s.d., $0.073$), while expanding the parameter count from 841,962 to 3,215,610. In the depth analysis, 2--8 layers produced accuracies of 0.925--0.928, whereas ten layers yielded the highest observed accuracy ($0.936\pm0.054$). Depths of 12--16 layers yielded accuracies of 0.921--0.930, while the primary eight-layer setting remained within 0.8 percentage points of the observed maximum. When graph-convolution order was varied, $K=4$ yielded $0.928\pm0.055$ accuracy, compared with $0.923\pm0.077$ at $K=2$ and 0.916 at $K=8$.

\Needspace{5\baselineskip}
\subsection{SEED-derived spatial and spectral signature}
\subsubsection{Spatial organization and channel pruning}

\begin{figure*}[htbp]
      \centering
      \includegraphics[width=\textwidth]{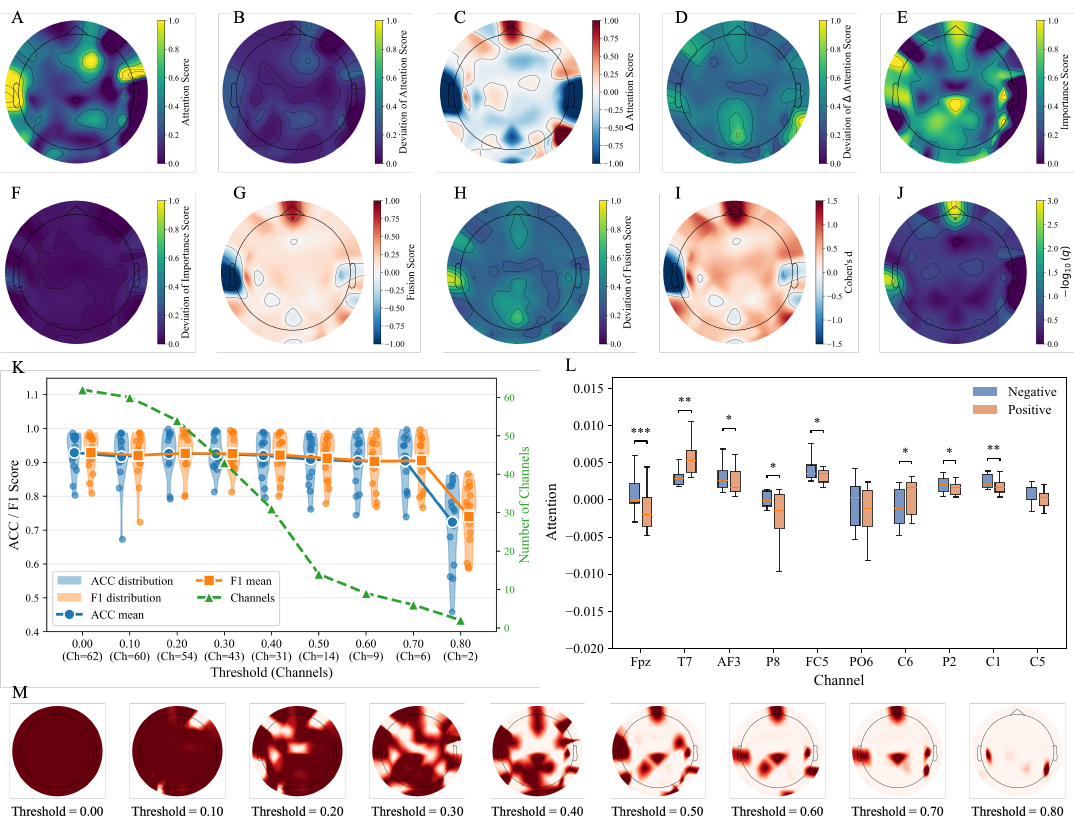}
      \caption{\textbf{Spatial readouts and channel pruning on SEED ($n=15$ participants).} Spatial condition contrasts are negative minus positive (class 0 minus class 1), with aggregation and scaling defined in Section~\ref{subsec:interpretability}. \textbf{A,} Scaled product of branch-mean attention. \textbf{B,} Across-participant standard deviation of the within-participant counterpart. \textbf{C,} Scaled pooled condition contrast. \textbf{D,} Sample-level standard deviation of scaled contrast vectors. \textbf{E,F,} Mean and across-participant standard deviation of normalized, fold-averaged mask scores. \textbf{G,H,} Mean and across-participant standard deviation of scaled mask-weighted contrasts. \textbf{I,} Standardized mean of the scaled scores. \textbf{J,} $-\log_{10}(q)$ from two-sided one-sample $t$-tests against the scaled zero, with Benjamini--Hochberg correction over 62 channels, where zero represents the midpoint of each participant's channel-score range. \textbf{K,} Participant-specific accuracy and positive-emotion F1 after training-fold channel selection and LOSO retraining, with points denoting test participants, lines denoting means and the green curve showing retained-channel counts averaged over LOSO folds. \textbf{L,} Raw participant-mean fused attention at the ten channels selected by \textbf{J} (blue, negative; orange, positive), with boxes showing medians and interquartile ranges and whiskers spanning the 5th--95th percentiles. Exploratory two-sided paired $t$-tests used Benjamini--Hochberg correction across these ten channels, which were selected and tested within the same cohort; *$q<0.05$, **$q<0.01$, ***$q<0.001$. \textbf{M,} Example topographies across pruning thresholds.}
      \label{fig:space_biomarker}
\end{figure*}

Fused attention, mask scores and their scaled combinations showed non-uniform channel profiles (Fig.~\ref{fig:space_biomarker}A--H), with the scaled mask-weighted score highlighting a distributed frontal, temporal, central and parietal pattern (Fig.~\ref{fig:space_biomarker}G). Within this pattern, Fpz was the sole channel for which the scaled score differed from zero after correction across 62 channels ($q=0.0038$; Fig.~\ref{fig:space_biomarker}J).

Exploratory paired comparisons of raw participant-level fused attention identified condition differences at eight of the ten selected channels (Fig.~\ref{fig:space_biomarker}L), comprising Fpz, T7, AF3, P8, FC5, C6, P2 and C1, whereas PO6 and C5 had corrected $q$ values at or above the significance threshold.

With channels selected using the training participants in each outer fold, increasing the pruning threshold reduced the mean retained-channel count across LOSO folds through 62, 60, 54, 43, 31, 14, 9, 6 and 2 (Fig.~\ref{fig:space_biomarker}K,M). Accuracy and F1 remained close to the full-montage values when an average of six channels was retained, whereas both metrics declined with an average of two channels.

\subsubsection{Frequency dependence of the SEED signature}

\begin{figure*}[htbp]
      \centering
      \includegraphics[width=\textwidth]{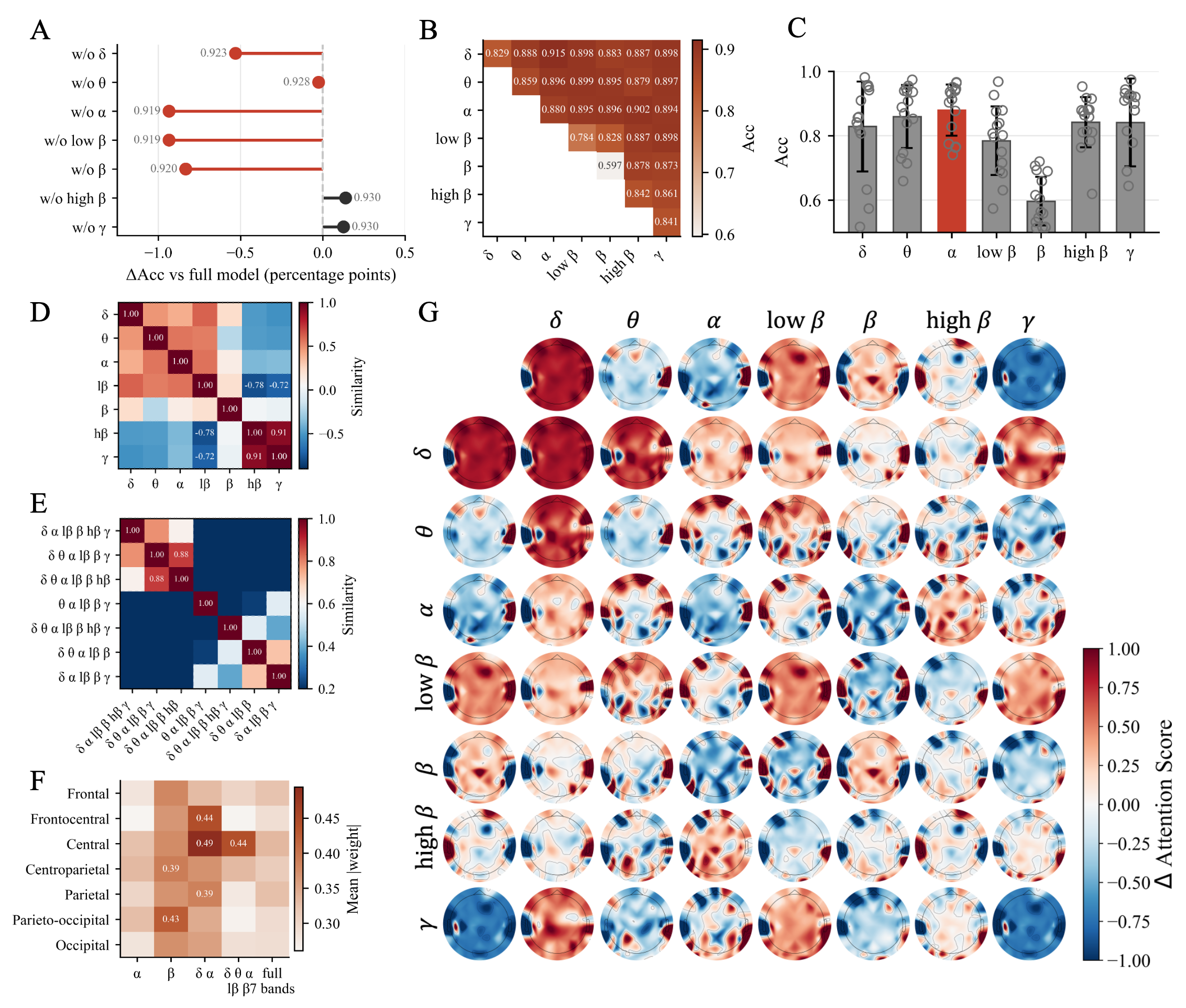}
      \caption{\textbf{Frequency dependence and spatial organization of the SEED-derived signature.} \textbf{A,} Accuracy of the full seven-band and leave-one-band-out models. \textbf{B,} Accuracy of single- and two-band inputs. \textbf{C,} Participant-level single-band accuracy ($n=15$; bars, means; error bars, s.d.). \textbf{D,} Correlations between spatial attention-contrast maps from single-band models. \textbf{E,} Similarity among representative multiband models. \textbf{F,} Regional mean absolute attention contrasts for representative models. \textbf{G,} Group-mean positive-minus-negative attention-contrast maps across bands and representative combinations.}
      \label{fig:freq_biomarker}
\end{figure*}

The full seven-band model achieved a mean accuracy of 0.928, with individual-band removal changing accuracy by $-0.9$ to $+0.2$ percentage points (Fig.~\ref{fig:freq_biomarker}A). The largest decreases followed removal of $\alpha$ or low $\beta$ (accuracy, 0.919), whereas removing high $\beta$ or $\gamma$ yielded an accuracy of 0.930. Among single bands, $\alpha$ produced the highest mean accuracy (0.880), whereas $\beta$ produced the lowest (0.597); the best two-band input, $\delta+\alpha$, reached 0.915, within 1.3 percentage points of the full model (Fig.~\ref{fig:freq_biomarker}B,C).

Spatial organization also changed with the spectral inputs supplied to the model; the high $\beta$ and $\gamma$ maps showed a strong correlation ($r=0.91$; Fig.~\ref{fig:freq_biomarker}D--G). Low $\beta$ correlated negatively with high $\beta$ ($r=-0.78$) and $\gamma$ ($r=-0.72$), while reduced multiband models differed in central and centroparietal weighting (Fig.~\ref{fig:freq_biomarker}E--G).
\Needspace{4\baselineskip}
\section{Discussion}

Across the public benchmarks, EmoDiPyraTrans achieved the highest mean accuracy and F1 among the evaluated methods, while the clinical transfer experiments characterized sensitivity to population shift and the SEED analyses identified spatial--spectral features associated with prediction. The magnitude and variability of these gains provide context for the benchmark rankings, which combine published SEED and FACED baseline values with rerun comparisons on the other public datasets. F1 improvements on FACED and DEAP remained below one percentage point, while DREAMER accuracy of 0.671 and an F1 of $0.328\pm0.273$ indicated substantial metric-specific and between-participant variation. Participant-level paired comparisons, balanced accuracy and recall would refine assessment of these small gains and clarify the DREAMER metric divergence by characterizing performance across classes and individuals.

The DEP-EEG comparisons extend this assessment to within-population decoding and transfer between clinically distinct groups \cite{huang2021neurofeedback,guan2025method}, with EmoDiPyraTrans achieving the leading HC$\rightarrow$DEP means. Transfer accuracy was 11.3 percentage points below DEP-LOSO, while mixed-population development also yielded lower accuracy and F1 than DEP-LOSO, highlighting the relevance of population composition to emotion decoding. Because cohort size and clinical heterogeneity varied together across protocols, the observed differences motivate further examination of population imbalance, negative transfer and internal partitioning as plausible contributors. Matched-size healthy-control subsets, population-balanced sampling and paired analyses over identical outer folds could distinguish these effects, while class-wise recall and macro-F1 would characterize variation in positive-emotion recognition across participants.

Within the model, the ablations support differential attention through consistent gains across both metrics and MMD through consistent gains in accuracy, whereas graph refinement and pseudo-label learning contributed according to the dataset and metric. Variation in graph-refinement effects may be associated with montage and cohort differences, while joint SEED ablations suggest an interaction between PL and MMD that warrants targeted analysis. These component-specific patterns motivate calibration-aware confidence selection or validation-based loss weighting \cite{arazo2020pseudo}, through which the contribution of PL could be adjusted to the target setting.

Complementing these performance comparisons, the spatial analyses relate relative channel weighting to raw condition differences, with each readout providing a distinct view of the model's spatial organization. Because channel-wise scaling places zero at each participant's score-range midpoint, the signs in Fig.~\ref{fig:space_biomarker}G locate channels relative to that midpoint, whereas panel L provides exploratory condition comparisons using raw attention at channels selected within the same cohort. Together, these scalp-level readouts generate spatial hypotheses that can be examined through source-resolved physiological measurements \cite{kappenman2010effects,antonakakis2020inter}, connecting model interpretation to the evaluation of underlying neural activity.

The associated SEED pruning results support compact predictive montages, as near-full performance was retained with an average of six channels under outer-fold evaluation. Whereas mask scores identify channels that sustain a fixed predictor, pruning evaluates performance after retraining, providing complementary evidence for the computational sufficiency of reduced channel sets. Replication in independent cohorts, together with comparisons against random and spatially matched montages, would assess how consistently this selection strategy preserves predictive performance across settings.

Alongside this spatial organization, the alpha-centred preference characterizes the spectral dependence of the SEED-derived signature, while the small effects of individual-band removal suggest that predictive information is distributed across the spectral representation. Configuration-dependent topographies further relate spatial interpretation to the spectral inputs supplied to the model, providing a basis for comparing channel weighting across frequency combinations. These patterns motivate hypotheses involving sensory gating, attention or affect regulation \cite{zheng2015investigating,luther2023oscillatory}, whose physiological relevance can be examined through independent measurements and experiments.

Further validation can strengthen assessment of shared physiological patterns by harmonizing dataset-specific montages, elicitation paradigms and label rules, while replicating the SEED signature derived from 15 participants across sessions, cohorts, clinical populations, tasks and hardware. The present evidence supports emotion decoding and computational sufficiency, providing a foundation for dedicated tests of neural necessity and diagnostic validity in physiological and clinical research. Extending the rPSD representation with phase, connectivity, cross-frequency coupling and event-related morphology would broaden the information available to the model, while repeated-seed evaluation would quantify optimization variability around the fixed-seed results. In parallel, external validation across additional settings would complement the local sensitivity scans by assessing the robustness of the framework under broader changes in data acquisition and population composition.

\section{Conclusion}
This study integrates cross-subject EEG emotion decoding with channel- and frequency-resolved model analysis through EmoDiPyraTrans, which achieved the highest participant-mean accuracy and positive-class F1 among the evaluated methods across five public benchmarks. Complementary DEP-EEG comparisons characterized population-transfer performance, with lower accuracy under healthy-to-depression transfer and mixed-population development than under within-population evaluation. On SEED, a distributed spatial pattern and an alpha-centred spectral preference were identified alongside configurations averaging six channels that preserved near-full performance, linking spatial and spectral patterns to predictive performance. Together, these results provide a framework in which cross-subject prediction and population-transfer performance are evaluated alongside model-derived candidate signatures, linking emotion decoding to the spatial and spectral information supporting prediction.

\bibliographystyle{IEEEtran}
\bibliography{ref}
\end{document}